\documentclass[aps,pre,floats,twocolumn,twoside,floatfix,superscriptaddress]{revtex4}
\usepackage{graphicx}
\usepackage{amssymb,amsmath}
\usepackage[utf8]{inputenc}

\newcommand{\rhoS}{\rho_{\scriptscriptstyle \rm S}}
\newcommand{\rhoE}{\rho_{\scriptscriptstyle \rm E}}
\newcommand{\rhoZ}{\rho_{\scriptscriptstyle \rm Z}}
\newcommand{\rhoR}{\rho_{\scriptscriptstyle \rm R}}
\newcommand{\rhoSE}{\rho_{\scriptscriptstyle \rm SE}}
\newcommand{\rhoSZ}{\rho_{\scriptscriptstyle \rm SZ}}
\newcommand{\rhoSR}{\rho_{\scriptscriptstyle \rm SR}}

\newcommand{\rhoZR}{\rho_{\scriptscriptstyle \rm ZR}}
\newcommand{\rhoSz}{\rho_{\scriptscriptstyle \rm S}^{\scriptscriptstyle (0)}}
\newcommand{\rhoEz}{\rho_{\scriptscriptstyle \rm E}^{\scriptscriptstyle (0)}}

\begin{document}
\title{Skepticism and rumor spreading: the role of spatial correlations}
\author{Marco Antonio Amaral}
\email{marcoantonio.amaral@cpf.ufsb.edu.br}
\affiliation{Instituto de Humanidades, Artes e Ciências, Universidade Federal do Sul da Bahia, CEP 45638-000, Teixeira de Freitas - BA, Brazil} 
\author{W. G. Dantas}
\affiliation{Departamento de Ciências Exatas, EEIMVR, Universidade Federal Fluminense, CEP 27255-125, Volta Redonda - RJ,
Brazil} 
\affiliation{Instituto de Física, Universidade Federal do Rio Grande do Sul, CEP 91501-970, Porto Alegre - RS, Brazil} 
\email{wgdantas@id.uff.br}
\author{Jeferson J. Arenzon}
\email{arenzon@ufrgs.br}
\affiliation{Instituto de Física, Universidade Federal do Rio Grande do Sul, CEP 91501-970, Porto Alegre - RS, Brazil} 
\affiliation{Instituto Nacional de Ciência e Tecnologia - Sistemas Complexos, Rio de Janeiro RJ, Brazil}

\date{\today}

\begin{abstract}
Critical thinking and skepticism are fundamental mechanisms that one may use to prevent the
spreading of rumors, fake-news and misinformation. We consider a simple model in which agents without
previous contact with the rumor, being skeptically oriented, may convince spreaders to stop their activity
or, once exposed to the rumor, decide not to propagate it as a consequence, for example, of fact-checking. 
We extend a previous, mean-field analysis of the combined effect of these two mechanisms, active and passive 
skepticism, to include spatial correlations. This can be done either analytically, through the pair
approximation, or simulating an agent-based version on diverse networks. Our results show that while in
mean-field there is no coexistence between spreaders and susceptibles (although, depending on the parameters,
there may be bistability depending on the initial conditions), when spatial correlations are included,
because of the protective effect of the isolation provided by removed agents,  coexistence is possible. 
\end{abstract}


\maketitle

\section{Introduction}
\label{intro}
Polarization and fake-news propagation are intertwined social phenomena that may be 
amplified by a systemic lack of critical thinking and skeptical skills among the population~\cite{BeCoDaScCaQu15,ViBeZoPeScCaStQu16,OcWe18,Baumann2020}.
Rumor or opinion spreading models~\cite{Galam08,CaFoLo09} provide simple descriptions of how misinformation and gossips diffuse
between networked individuals, in a similar way as contagious diseases propagate through direct contact~\cite{Britton10,PaCaMiVe15,Sun16,ArRoMo18}. 
In the later, where the contamination is usually involuntary and unconscious, a deeper understanding of the process has led to a more effective protection
of susceptible populations through better designed vaccination~\cite{MoWi74,Strassburg82,FuSaJa10,Smith2012,Wang2016,BoBoMaGr19,CoFe20}, isolation~\cite{YaTaWa18,BiGo20} and quarantine~\cite{Kato2011,DoGaRaBa13,HaNe17} campaigns. The comprehension of
the main underlying mechanisms in the case of rumor propagation, involving an intentional and directed action from the spreader, may help to devise efficient strategies to 
counteract the effects of anti and pseudoscience movements. Similar programs have been
proposed, for example, to confront internet hate speech~\cite{Johnson19}, organized 
crime~\cite{CuGo18}, and terrorism~\cite{GaJa16,Santoprete19}.

Many analogies have been drawn between the processes of rumor and disease propagation through contact. 
The simplest models for the above contact processes~\cite{MaDi99} explore their similarity and consider
three dynamical states: ignorants, spreaders and stiflers~\cite{GoNe64,DaKe64,DaKe65,GoNe67,MaTh73} for rumor
propagation and susceptible, infected and removed~\cite{KeMc27} for diseases. 
While unaware of the rumor or uncontaminated by the disease, the agent is susceptible (S) to it. 
After being exposed, if the agent does not become a spreader (state Z), it turns into a stifler (for
example, resisting and getting removed from the propagation pool, state R).
More refined models introduced 
further states and mechanisms~\cite{CaFoLo09,PaCaMiVe15,Brauer17}. For example, there is an
intermediate, exposed state -- that is relevant for the model introduced here --
corresponding to a non contagious period after the contact with a spreader~\cite{ArSc84,XiJiSoSo15}, when the
agent hesitates between becoming Z or R. 

Fake-news, misinformation, hoaxes and pseudosciences propagate among the population through least
resistance paths, getting reinforced by new technologies, and are rarely completely removed. Several works have tried to measure, identify 
and model their mechanisms and effects~\cite{TaOlCiRu18,ZeGa19}. An essential ingredient is the presence of skeptical agents.
In standard
disease or rumor propagation, only the susceptible is affected by its interaction with
a spreader. Skepticism may, instead, not only prevent the susceptible to become infected
but also i) change the spreader, removing it from the propagation pool and ii) prevent an
already exposed agent to become a spreader. While an active skeptical susceptible may
convince its neighbor spreader, the convincing process of an exposed individual is
passive (by consulting the original sources and the available literature, for example).  
In a previous work~\cite{AmAr18}, a simple model introduced the opposing behavior of
skeptical agents and was studied within the one-site mean-field approximation. However,
the effect of spatial correlations was not taken into account. Here we extend the
analysis of Ref.~\cite{AmAr18} and consider results for i) the pair-approximation that
improves on the mean-field analysis and ii)  an agent-based version on 
one- and two-dimensional regular lattices and complex networks (random and scale-free).
Interestingly, there is a similarity between the behavior of skeptical agents in our
model~\cite{AmAr18} and the mechanism introduced in the context of a
zombie outbreak~\cite{MuHuImSm09,AlBiMySe15,HoWa20}, whose analogy with real pandemics
have been previously explored for didactic purposes~\cite{CDC,VeCrCaJaAm14}. 

The paper is organized as follows. Section~\ref{section.model} describes our model 
and its dynamical rules. Section~\ref{section.MF} summarizes the mean-field results
of Ref.~\cite{AmAr18} and introduces the improved equations within the pair
approximation, along with its numerical solutions in both 1D and 2D. We then explore,
with numerical simulations (and compare with the results of the previous section), the 
behavior of the model when spatial correlations are taken into account, both in
regular (1D and 2D), Section~\ref{section.regular}, and complex networks (random and
scale-free), Section~\ref{section.complex}. Finally, the discussion and conclusions
are presented in Section~\ref{section.conclusion}.

\section{The model} 
\label{section.model}

Rumor spreading is modeled here with a set of four possible states (compartments). 
Some individuals, labeled $Z$, attempt to propagate their opinion  by convincing a neighbor 
who had no previous contact with the information being transmitted. The neighbor is thus susceptible, albeit possibly skeptical (either way, we call it $S$).
There are two possible outcomes for such pairwise, catalytic interaction between $S$ and $Z$,
depending on which agent is modified. First, the $S$ agent may  
get exposed ($E$) to the information carried by $Z$ with 
probability $\beta$:
\begin{equation}
SZ\stackrel{\beta}\longrightarrow EZ. 
\label{eq.beta}
\end{equation}
This is similar to the standard SIR model with exposed agents (SEIR).
Another possible result of this interaction, that depends on the degree of (active) skepticism, is the spreader $Z$, after
being convinced by a skeptical $S$, gets removed ($R$) from the population of propagators with 
probability $\kappa$:
\begin{equation}
  ZS\stackrel{\kappa}\longrightarrow RS.
\label{eq.kappa}
\end{equation}
This later process, where the rumor propagation can be deflected, is not usually 
possible in disease propagation.
Indeed, in standard SIR models, the spreader does not get modified by interacting with a susceptible. While in the exposed, latent state, the agent may get effectively convinced and, spontaneously, become a spreader with
probability $\gamma$,
\begin{equation}
  E\stackrel{\gamma}\longrightarrow Z.
  \label{eq.gamma}
\end{equation}  
Otherwise, during the $E$ state, the (skeptical) agent may check the information received and doubt it.  
As a consequence, it may get removed without any external interaction (passive skepticism), with probability $1-\gamma$:
\begin{equation}
  E\stackrel{1-\gamma}\longrightarrow R.
  \label{eq.1mgamma}
\end{equation}  
By pulling itself out of the spreading process, the agent is no longer capable of 
changing its state, becoming immune to any further contact with the rumor 
but also taking no action against the spreaders. 

The two possible outcomes discussed above, for the $SZ$ interaction, as a consequence of skeptical 
and critical thinking, justifies the analogy developed in Ref.~\cite{AmAr18} between this 
bidirectional mechanism with the pop-culture concept of zombies and the related apocalypse.
Interestingly, several mathematical models have been studied considering such apocalyptic 
scenarios~\cite{MuHuImSm09,AlBiMySe15,HoWa20}. Our model has an analogous mathematical structure albeit its interpretation does not
deal with the living-dead but, instead, with the current crisis where fake-news and
disinformation may be out of control with profound consequences.

\section{Mean field and pair approximation}
\label{section.MF}

The density of each compartment evolves in time, considering Eqs.~(\ref{eq.beta})-(\ref{eq.1mgamma}), as~\cite{AmAr18}
\begin{align}
\begin{aligned}
  \dot{\rho}_{\scriptscriptstyle \rm S}&=-\beta\rhoSZ\\
  \dot{\rho}_{\scriptscriptstyle \rm E}&=\beta\rhoSZ -\rhoE\\
  \dot{\rho}_{\scriptscriptstyle \rm Z}&=\gamma \rhoE-\kappa\rhoSZ\\
  \dot{\rho}_{\scriptscriptstyle \rm R}&=(1-\gamma)\rhoE+\kappa\rhoSZ.
\end{aligned}
\label{eq.onesite}
\end{align}
There are two integrals of motion, $\rhoS+\rhoE+\rhoZ+\rhoR=1$ and
$P\equiv (\beta\gamma-\kappa)\rhoS+\beta\gamma\rhoE+\beta\rhoZ$, reducing the
number of independent variables. 
Within the one-site, mean-field (MF) approximation that neglects correlations between different sites, $\rhoSZ=\rhoS\rhoZ$, 
there are  two possible  steady states~\cite{AmAr18}, $(\rhoS^*,\rhoE^*,\rhoZ^*,\rhoR^*)$: 
either the spreaders are absent in the population, $(\rhoS^*,0,0,1-\rhoS^*)\equiv F_{\scriptscriptstyle \rm S}$, or the susceptibles, 
$(0,0,\rhoZ^*,1-\rhoZ^*)\equiv F_{\scriptscriptstyle \rm Z}$.
The stability of these solutions depends on 
$\beta,\gamma$ and $\kappa$, as well as on the initial condition for the densities,
chosen to consist entirely of susceptible and exposed individuals, $\rhoSz+\rhoEz=1$. 
With this choice, $F_{\scriptscriptstyle \rm S}$ is stable whenever $\beta\gamma/\kappa<1$ and 
$\rhoSz>\beta\gamma/\kappa$. Otherwise, $F_{\scriptscriptstyle\rm Z}$ is the stable solution. Thus, at the MF level, there
is no stationary states where spreaders and susceptibles coexist in a polarized state. 
The rumor either spreads over a maximum number of agents 
($F_{\scriptscriptstyle \rm Z}$ solution) or vanishes, with part of the population remaining unexposed 
($F_{\scriptscriptstyle \rm S}$ solution). However, by including spatial correlations,
even at the level of pair approximation (PA)~\cite{MaDi99}, this scenario changes, as shown below.

We extend the analysis of Ref.~\cite{AmAr18} to include correlations between pairs 
of sites~\cite{MaDi99}, that are particularly important in models with catalytic reactions. 
The time evolution of the densities of pairs of strategies
now depends on the probability of having triplets: 
\begin{align}
\begin{aligned}
  \dot{\rho}_{\scriptscriptstyle\rm SS} &=-\beta\rho_{\scriptscriptstyle\rm SSZ} \\
  \dot{\rho}_{\scriptscriptstyle\rm EE}&=-2\rho_{\scriptscriptstyle\rm EE}+\beta\rho_{\scriptscriptstyle\rm ESZ}\\
  \dot{\rho}_{\scriptscriptstyle\rm ZZ}&=2\gamma\rho_{\scriptscriptstyle\rm EZ}-\kappa\rho_{\scriptscriptstyle\rm SZZ}\\
  \dot{\rho}_{\scriptscriptstyle\rm RR}&=2(1-\gamma)\rho_{\scriptscriptstyle\rm ER}+\kappa\rho_{\scriptscriptstyle\rm SZR}\\
   \dot{\rho}_{\scriptscriptstyle\rm SE} &= -\rhoSE+ \beta\left(1-\frac{1}{2d}\right)(\rho_{\scriptscriptstyle\rm SSZ}-\rho_{\scriptscriptstyle\rm ZSE})\\
  \dot{\rho}_{\scriptscriptstyle\rm SZ} &= \gamma\rhoSE -\frac{\beta+\kappa}{2d}\rhoSZ- \left(1-\frac{1}{2d}\right) (\beta\rho_{\scriptscriptstyle\rm ZSZ}+\kappa\rho_{\scriptscriptstyle\rm SZS})\\
 \dot{\rho}_{\scriptscriptstyle\rm SR} &= (1-\gamma) \rhoSE+\frac{\kappa}{2d} \rhoSZ-\left(1-\frac{1}{2d}\right) (\beta\rho_{\scriptscriptstyle\rm ZSR}-\kappa \rho_{\scriptscriptstyle\rm SZS})\\
  \dot{\rho}_{\scriptscriptstyle\rm ZE} &= \gamma\rho_{\scriptscriptstyle\rm EE} -\rho_{\scriptscriptstyle\rm EZ}+ \frac{\beta}{2d} \rhoSZ+ \left(1-\frac{1}{2d}\right)
(\beta\rho_{\scriptscriptstyle\rm ZSZ}  -\kappa\rho_{\scriptscriptstyle\rm SZE})\\
  \dot{\rho}_{\scriptscriptstyle\rm ER} &=(1-\gamma) \rho_{\scriptscriptstyle\rm EE}- \rho_{\scriptscriptstyle\rm ER}+ \left(1-\frac{1}{2d}\right) (\beta\rho_{\scriptscriptstyle\rm ZSR}+\kappa \rho_{\scriptscriptstyle\rm SZE})  \\
   \dot{\rho}_{\scriptscriptstyle\rm ZR} &= \gamma\rho_{\scriptscriptstyle\rm ER}+(1-\gamma)\rho_{\scriptscriptstyle\rm EZ}+\kappa \left(1-\frac{1}{2d}\right) (\rho_{\scriptscriptstyle\rm SZZ}-\rho_{\scriptscriptstyle\rm SZR}).
\end{aligned}
\label{eq.pair}
\end{align}
The number of equations can be further reduced by employing sum rules like
$\rhoS=\sum_x\rho_{{\scriptscriptstyle\rm S}x}$ and 
$\rhoSZ=\sum_x\rho_{{\scriptscriptstyle\rm SZ}x}$. With the pair approximation,
$\rho_{xyz} \simeq \rho_{xy}\rho_{yz}/\rho_{y}$, the above equations can be closed. All stationary solutions have $\rhoSZ=0$ (i.e., no
reacting pairs) and $\rho_{{\scriptscriptstyle\rm E}x}=0$ because exposed agents eventually decay, $\rhoE=0$. However, since no condition applies to $\rhoSR$ and $\rhoZR$, besides the mean-field $F_{\scriptscriptstyle\rm S}$ and $F_{\scriptscriptstyle\rm Z}$ solutions~\cite{AmAr18}, susceptible and spreaders may coexist, but not interact, i.e., their interactions are screened by removed individuals. Differently from the single-site, MF approximation~\cite{AmAr18}, we were not able to obtain an analytical expression for the densities at the stationary state. We resort, instead, to numerical methods, integrating Eqs.~(\ref{eq.pair}) with a fourth-order Runge-Kutta method. Notice that, in MF, for a given probability $\gamma$ of an exposed agent becoming a spreader, an increase in the exposition probability ($\beta$) could be compensated by an equivalent increase in the active skepticism ($\kappa$), i.e., the results only depend on $\beta\gamma/\kappa$. The PA behavior, on the other hand, do not depend on such a simple ratio.  Thus, we consider three particular cases, albeit representative of the overall behavior, in order to compare how spreaders and susceptibles compete. For $\beta\ll\kappa$, the population skepticism toward the rumor is much stronger than the rumor virality, while in the other extreme, $\beta\gg\kappa$, it is the opposite and the rumor outcompetes the skepticism. We also consider the intermediate case, $\beta=\kappa$, where both tendencies are similar. 


\begin{figure}[htb!]
\begin{center}
\includegraphics[width=\columnwidth]{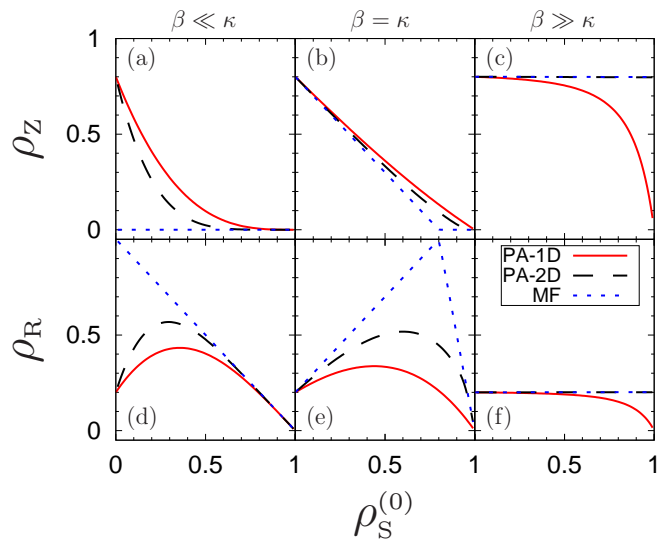}
\caption{Density of spreaders $\rhoZ$ (top row) and removed $\rhoR$ (bottom row), as 
a function of the initial fraction of susceptibles, $\rhoSz=1-\rhoEz$. The results
allow the comparison between the different levels of approximation: mean-field  (dotted
line) and the pair approximation for 1D (solid line) and 2D (dashed line). The relation between
$\beta$ and $\kappa$ is indicated in each column, from left to right: $\beta\ll\kappa$ ($\beta=10^{-3}$ and $\kappa=0.8$),
$\beta=\kappa=0.1$ and $\beta\gg\kappa$ ($\beta=0.8$ and $\kappa=10^{-3}$). Notice that in panels (c)
and (f), the MF and the PA for 2D coincide. In all cases, $\gamma=0.8$.}
\label{fig.density-ZR_vs_S0}
\end{center}
\end{figure}

Fig.~\ref{fig.density-ZR_vs_S0} compares the asymptotic fraction of spreaders and removed individuals as a function
of the initial number of susceptibles, $\rhoSz$, for the PA in 1D and 2D, with the MF results. 
We consider a large probability of transforming  exposed individuals into 
spreaders ($\gamma=0.8$), and several ratios $\beta/\kappa$.
For $\rhoSz=0$, a fraction $\gamma$ of the all-exposed initial population turns, spontaneously, into spreaders, $\rhoZ\simeq\gamma$, the remaining becoming removed, $\rhoR\simeq 1-\gamma$. 
For small to moderate values of $\beta/\kappa$, spreaders only prevail when the number of susceptibles is small,  
Figs.~\ref{fig.density-ZR_vs_S0}a and b. 
As $\beta$ increases, facilitating the transition $S\to E$ (and, then, to $Z$), there is
a change in the concavity of the curve $\rhoZ(\rhoSz)$: the initial decreasing rate of $\rhoZ$
becomes very small (Figs.~\ref{fig.density-ZR_vs_S0}a to c) and the
overall fraction of spreaders gets larger, in both 1 and 2D,  despite $\rhoZ$ being always a monotonically decreasing 
function of $\rhoSz$.

For $\beta\gg\kappa$, Fig.~\ref{fig.density-ZR_vs_S0}c, the asymptotic fraction of spreaders may be large 
even for populations initially dominated by susceptibles. In this limit, the information is highly
viral and almost no spreader gets
removed after being created from the initially exposed individuals. Indeed, in 2D, almost independently of 
$\rhoSz$ (except, of course, at $\rhoSz=1$ where $\rhoZ=0$ and it is discontinuous), $\rhoZ\simeq\gamma$ and $\rhoR\simeq 1-\gamma$ (then, $\rhoS\simeq 0$). 
This constant behavior, present only in 2D, is remarkable since the more susceptibles (that may be skeptics) are present in the initial state, the more resistance is to be expected, helping to prevent the increase of spreaders, as observed in all other cases. 
There is, however, a qualitative difference with the 1D case where
there is an appreciable reduction in both $\rhoZ$ and $\rhoR$ when $\rhoSz$ increases (i.e., $\rhoS>0$). 
This strong dependence on the dimension has its origin in the different coordination and
the possibility of having multiple paths connecting two sites in 2D. In 1D, on the contrary,  
groups of susceptible 
agents may easily become isolated once they are located between two removed individuals, preventing their
exposition to spreaders and halting any further transformation. As a consequence,
$\rhoS$ remains finite and increases with $\rhoSz$, while $\rhoZ$ decreases. 
 

Fig.~\ref{fig.density-ZR_vs_S0} also shows (bottom row) the population of removed individuals, $\rhoR$, 
as a consequence of the skepticism among the population. Their number is relevant when designing a 
strategy to prevent rumor spreading.
Up to the level of the pair-approximation, in spite of the rumor never completely vanishing,
some control over its global spreading can be obtained by properly
tunning either the initial conditions or the parameters of the model. Specifically,
despite our results never showing an all-removed scenario, there usually is a
maximum value of $\rhoR(\rhoSz)$, that optimizes the resistance
against the rumor, Figs.~\ref{fig.density-ZR_vs_S0}d and e, for $\beta\leq\kappa$.
Nonetheless, for $\beta\gg\kappa$, Fig.~\ref{fig.density-ZR_vs_S0}f, the 
maximum $\rhoR$ always occurs at $\rhoSz=0$. 
On the other hand,  as a function of $\gamma$ (not shown), all cases present a maximum.

The main observed difference between the pair and mean-field
approximations is related to the possibility of spreaders, susceptibles 
and removed agents coexisting. Within the PA this is, in fact, the only solution 
found, even for the extremes $\beta\ll\kappa$ and $\beta\gg\kappa$, where the population 
of spreaders and susceptibles, respectively, becomes very small, but does not disappear. Indeed, 
a more extensive analysis would be necessary before ruling out solutions like 
$F_{\scriptscriptstyle\rm S}$ or $F_{\scriptscriptstyle\rm Z}$ that seem to appear, rather trivially,
for $\rhoSz=0$ and 1. However, these states also occur
when $\beta=1, \kappa=0$ with $\gamma\approx 0$, a particular situation
where the final population is formed only by susceptibles
and removed individuals ($F_{\scriptscriptstyle\rm S}$), or if $\beta=0, \kappa=1$ with
$\gamma\approx 1$, when the steady state is $F_{\scriptscriptstyle\rm Z}$. Both cases
are independent of the initial population of susceptibles, $\rhoSz$.


\section{Spatially distributed systems}
\label{section.spatial}

In the previous section we improved on the MF analysis of Ref.~\cite{AmAr18} 
by taking into account correlations on the level of couples of sites (the PA). These 
results should be compared with the simulations on spatial lattices, exploring the
effects of these correlations on the steady state. 
We consider three different networks on which $N$ sites are arranged: random, complex (scale-free)
and $d$-dimensional regular lattices with side $L$ ($N=L^d$) and periodic boundary conditions.  
In each site there is a single agent that, initially, is set to be either $E$ or $S$. 
Along the dynamics, a site is chosen at random and updated accordingly with
Eqs.~(\ref{eq.beta})-(\ref{eq.1mgamma}): an $E$ decays into $Z$ or $R$ with probability $\gamma$ or $1-\gamma$, respectively, irrespective of the neighborhood, while if it is a $Z$ or an $S$, a neighbor is also randomly chosen and, if appropriate, either Eq.~(\ref{eq.beta}) or (\ref{eq.kappa}) is used 
with the corresponding probability. The asymptotic densities are measured for different system sizes and averages are taken over
100 to 200 samples.
When the size is not explicitly mentioned, the results were extrapolated to $L\to\infty$ using a simple scaling, $\rho_x^{(\infty)}-\rho_x^{(L)}\sim L^{-1}$,
which fits quite well the numerical data.

\subsection{Regular lattices}
\label{section.regular}

\subsubsection{One-dimensional case}

\begin{figure}[htb!]
\begin{center}
\includegraphics[width=\columnwidth]{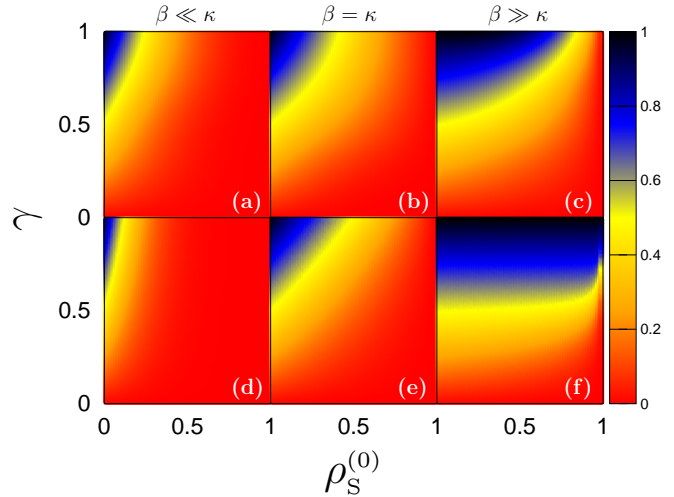}
\caption{Density of spreaders $\rhoZ$ for spatial systems in 1D ($L=5\times 10^5$, top row) and 2D ($L=500$, bottom row). 
As indicated on the top of each column, from left to right: $\beta\ll\kappa$, $\beta=\kappa$, and $\beta\gg\kappa$. In all
cases, $\gamma=0.8$.}
\label{fig.maps}
\end{center}
\end{figure}

\begin{figure}[htb!]
\begin{center}
  \includegraphics[width=\columnwidth]{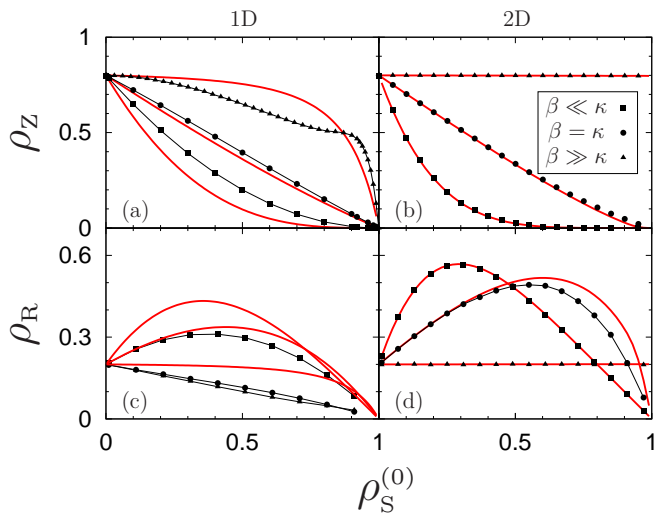}
\caption{Density of spreaders $\rhoZ$ (top) and removed (bottom) as a function of $\rhoSz$ for both 1D (left) and 2D (right). The
solid lines are the results from the PA with $\gamma=0.8$, while the points are extrapolations, for $L\to\infty$, of the corresponding simulations. Notice the excellent agreement in the 2D case.} 
\label{fig.regular-ZR_vs_S0}
\end{center}
\end{figure}

We consider linear systems with periodic boundary conditions and sizes ranging from $L=10^5$  to $10^6$.
The behavior of $\rhoZ$, shown in Fig.~\ref{fig.maps} for several values of the parameters,
is qualitatively similar to the PA. Spreaders and susceptibles coexist 
by forming isolated islands where the intermediate removed individuals prevent any interaction.
There are, however, some differences between both cases. In spatial systems, spreaders are the prevalent species in a slightly smaller 
region of the $(\rhoSz,\gamma)$ plane. Moreover, as can be seen in Fig.~\ref{fig.regular-ZR_vs_S0}a, the PA usually
underestimates $\rhoZ$ (see, for example, the curves for $\beta\ll\kappa$ and $\beta=\kappa$).
However, for $\beta\gg\kappa$, $\rhoZ$ develops a small plateau (observed only in the spatial 1D case), before which
the pair-approximation gives a larger estimate for $\rhoZ$. Although there is a reasonable quantitative agreement between the simulation
and the PA for $\rhoZ$, the comparison with $\rhoR$ is not as good, as can be seen in Fig.~\ref{fig.regular-ZR_vs_S0}c. 
Another interesting feature is that usually the partition between spreaders and susceptibles is 
asymmetric, even for $\beta=\kappa$, with the maximum symmetry (measured, for example, by the product $\rhoZ\rhoS$)
occurring for $\beta\gg\kappa$ in a domain where $\gamma\approx 1$ and $\rhoSz\approx 1$.

\subsubsection{Two-dimensional case}



In 2D, we consider square lattices with $L^2$ sites ($L=10^2-10^3$) and 
periodic boundary conditions along both directions. 
Differently from the 1D case, the agreement between the simulation
and the PA is excellent, suggesting that the reactions induce
strong pair correlations that dominate the stationary state. 
This short-range coupling is related to the compact clusters of
spreaders or susceptibles, whose interactions are inhibited by intermediate
removed individuals. These $R$ agents,
since they no longer interact, prevent any further evolution of the system. Models
that are similar, but that do not sustain an equilibrium coexistence of 
three species, do not show such agreement between the PA and
simulations.



The density of removed individuals measured in the simulation agrees well with 
the pair approximation predictions, Fig.~\ref{fig.regular-ZR_vs_S0}d, although
some deviation can be observed for $\beta=\kappa$ and large $\rhoSz$. 
Relevant information may be also obtained from
the position and height of the peak of $\rhoR$
that are, respectively, an increasing and decreasing  
function of $\gamma$ (not shown).
Thus, for a given
virality, there is an optimal fraction of susceptibles/skeptical agents that,
if present in the beginning of the propagation process, will counteract by
producing the largest fraction of removed individuals.
When $\gamma$ is large, exposed individuals are easily converted into spreaders
and, because this maximum value
increases with $\gamma$, the initial concentration of susceptibles has to be large enough 
to enforce a maximum removal due to skepticism.
Notice that even in the limit $\gamma\to 1$ (high virality), the amount of these removed agents is still large (not shown). 
Nonetheless, the effectiveness, i.e., 
the actual fraction of removed agents at the maximum,  
decreases with $\gamma$.  


\subsection{Scale-free and random networks}
\label{section.complex}

\begin{figure}[htb]
\begin{center}
\includegraphics[width=\columnwidth]{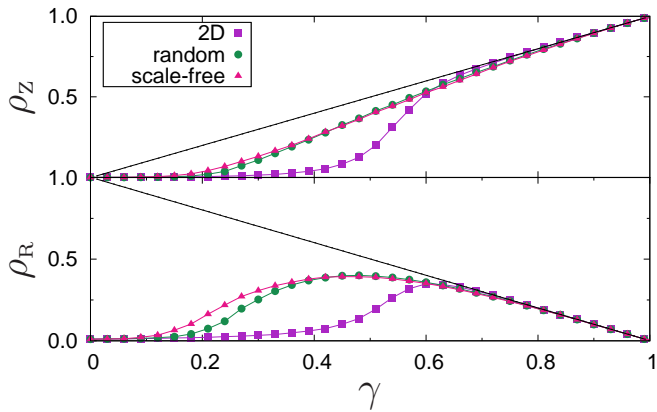}
\caption{Asymptotic fraction of spreaders ($\rhoZ$, top panel) and removed agents ($\rhoR$, bottom panel)  as a function of $\gamma$ with $N=10^4$ for regular, random and scale-free networks, along with
the mean-field result (straight line). We consider $\beta\gg\kappa$ and $\rhoSz=0.99$.} 
\label{fig.SZ-gamma}
\end{center}
\end{figure}

Social networks are neither fully-connected nor regular, being indeed
better described by complex networks. We thus extend the previous 
analysis to include random and scale-free networks, both built using the Krapivsky-Redner algorithm~\cite{KrRe01}. 
Fig.~\ref{fig.SZ-gamma} presents the behavior of $\rhoZ$ and $\rhoR$, as a function of $\gamma$, in the region 
$\beta\gg\kappa$ for the regular, random and scale-free networks, along with the MF approximation 
results (straight lines). In this regime of high rumor virality, 
all densities are much more sensitive to $\gamma$ than to $\rhoSz$. Except for large $\gamma$ where MF nicely agrees with
the simulations (and all three networks behave the same), both spreaders (top panel) and removed (bottom panel)
are overestimated in MF while susceptibles are underestimated. 

For the model we consider here, the overall qualitative behavior observed on these three structures is similar. Quantitatively, although the difference between the random and
the scale-free networks is small in the whole interval, they strongly differ from the square lattice where the amount of removed agents is much smaller. This robustness regarding the topology of connections seems to have a common origin on the strong pair correlations, as indicated by the good agreement with the pair-approximation results. Moreover, it is related to the permanent screening role of the removed, preventing further interactions between
susceptibles and spreaders and, consequently, halting the dynamics (exposed agents have a similar, albeit transient role). 
In order to be effective, any strategy cluster must be surrounded by removed individuals. This is the case of the square lattice since the nearest-neighbor groups with either $S$ or $Z$ agents are more compact and, because of that, need a smaller number of removed to fully cover their
surface. Once long-range connections are present, as in the random and scale-free networks, such surface becomes larger and more diffuse,
and a larger number of removed individuals is necessary to cover it. In the MF limit, where the network is fully-connected, it is no longer possible to
prevent the interactions between spreaders and susceptibles and they cannot coexist. Indeed, the asymptotic state becomes
a mixture of removed and either susceptibles ($F_{\scriptscriptstyle\rm S}$) or spreaders ($F_{\scriptscriptstyle\rm Z}$).

These results hint to a robust effect, where the sole nature of the interactions, either short or long-range,
determines whether the fraction of removed agents is enough to stabilize the coexistence state. While long range
interactions can initially lead to a wider spreading of the rumor, at the same time the $SZ$ interactions produce a larger fraction of removed individuals in the final state. This is observed (Fig.~\ref{fig.SZ-gamma}) both in random
and complex networks, where despite the distribution of connectivity being different, the results are much alike. The local
pattern of connections in these lattices differs from the square lattice, making the protective effect of removed agents less
efficient. 
Indeed, on a square lattice, because connected susceptibles form a compact group whose surface is much smaller, less removed 
individuals are necessary to cover such surface, protecting the susceptibles. The more long-range interactions are present, the more 
difficult it becomes to completely surround the group. Because of this, random and complex networks are more easily invaded by 
spreaders, leading to a larger number of removed agents. The fully connected limit corresponding to MF is an extreme case where such protection is impossible, preventing any coexistence between spreaders and susceptibles. 
On the square lattice, because the number of removed and spreaders is smaller than in complex networks, there are more
susceptibles. This is a general feature, observed on most of the parameter space. 

There are, in this model, two possible ways to control the propagation of a rumor and, consequently, have a small asymptotic fraction of spreaders, $\rhoZ \rightarrow 0$. A direct mechanism occurs when the passive skepticism within the population is large ($\gamma$ is small and almost no exposed individual becomes a spreader). Nonetheless, typical populations do not have a small $\gamma$. Usual, intermediate values
of $\gamma$ allow an indirect mechanism with a larger number of removed agents that protect the population of susceptibles, isolating them from the spreaders and allowing a stable coexistence between the two populations.

\begin{figure}[htb]
  \centering
  {\includegraphics[width=0.15\textwidth]{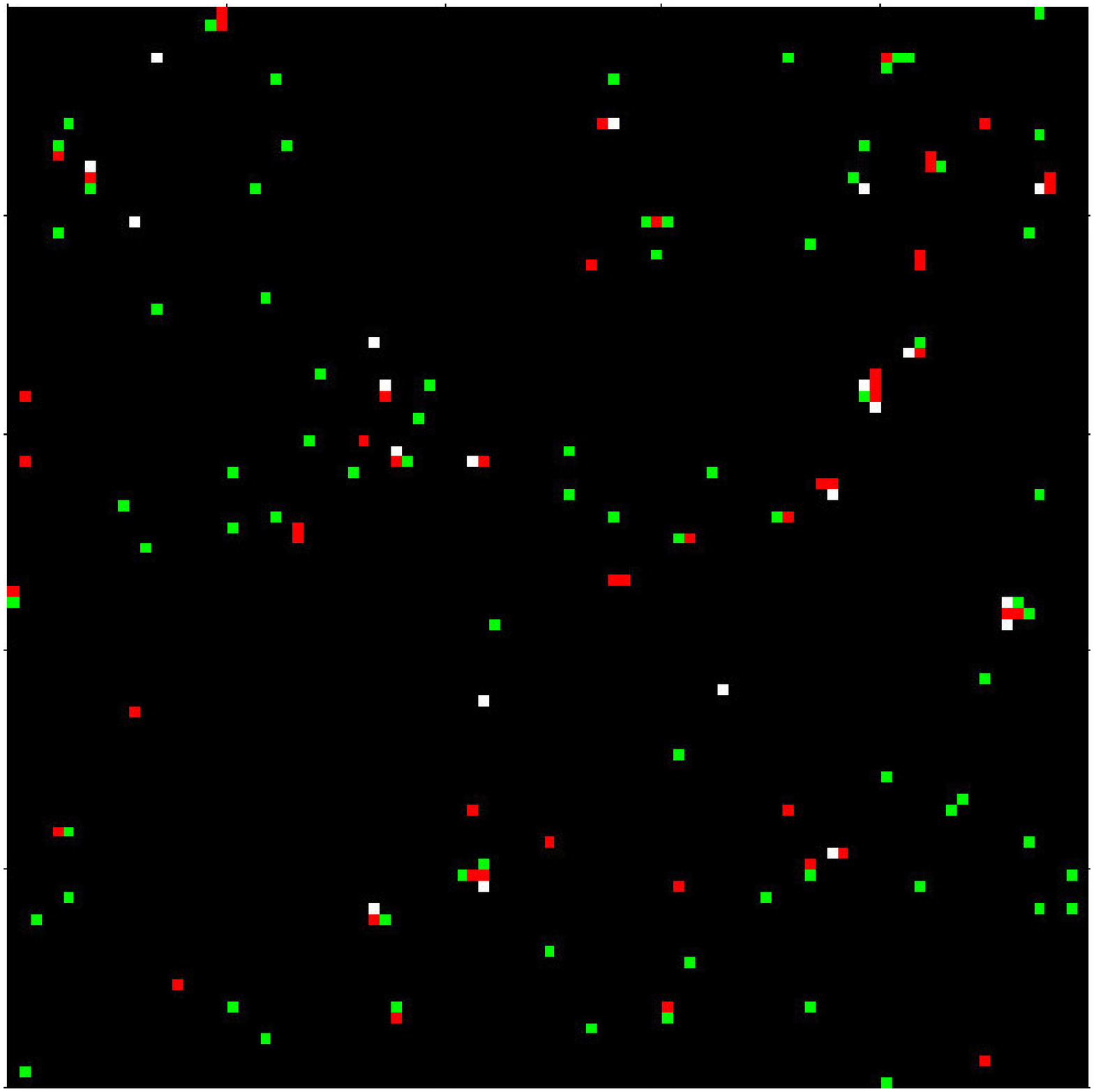}} \hspace{0.5mm}
  {\includegraphics[width=0.15\textwidth]{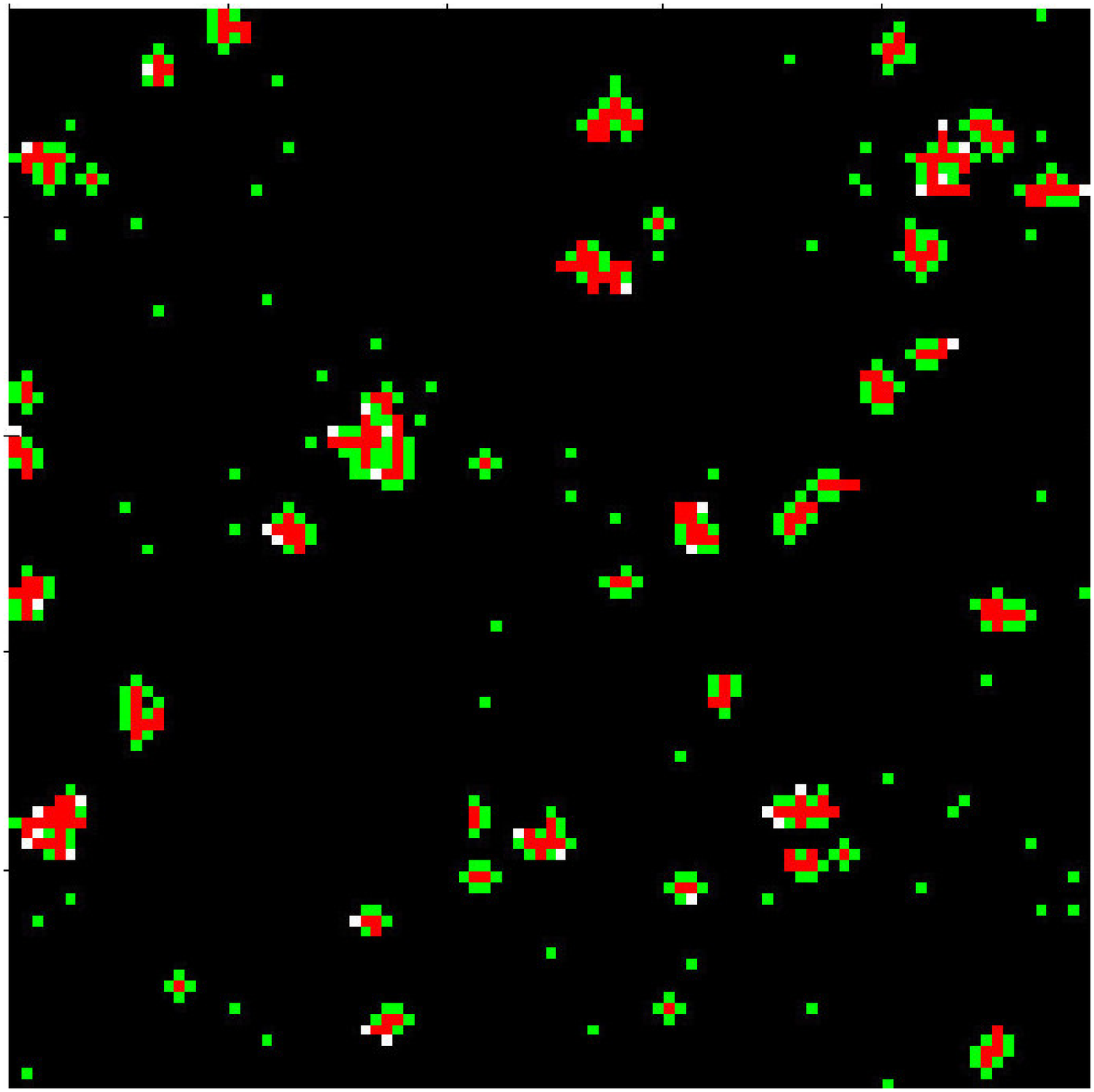}} \hspace{0.5mm}
  {\includegraphics[width=0.15\textwidth]{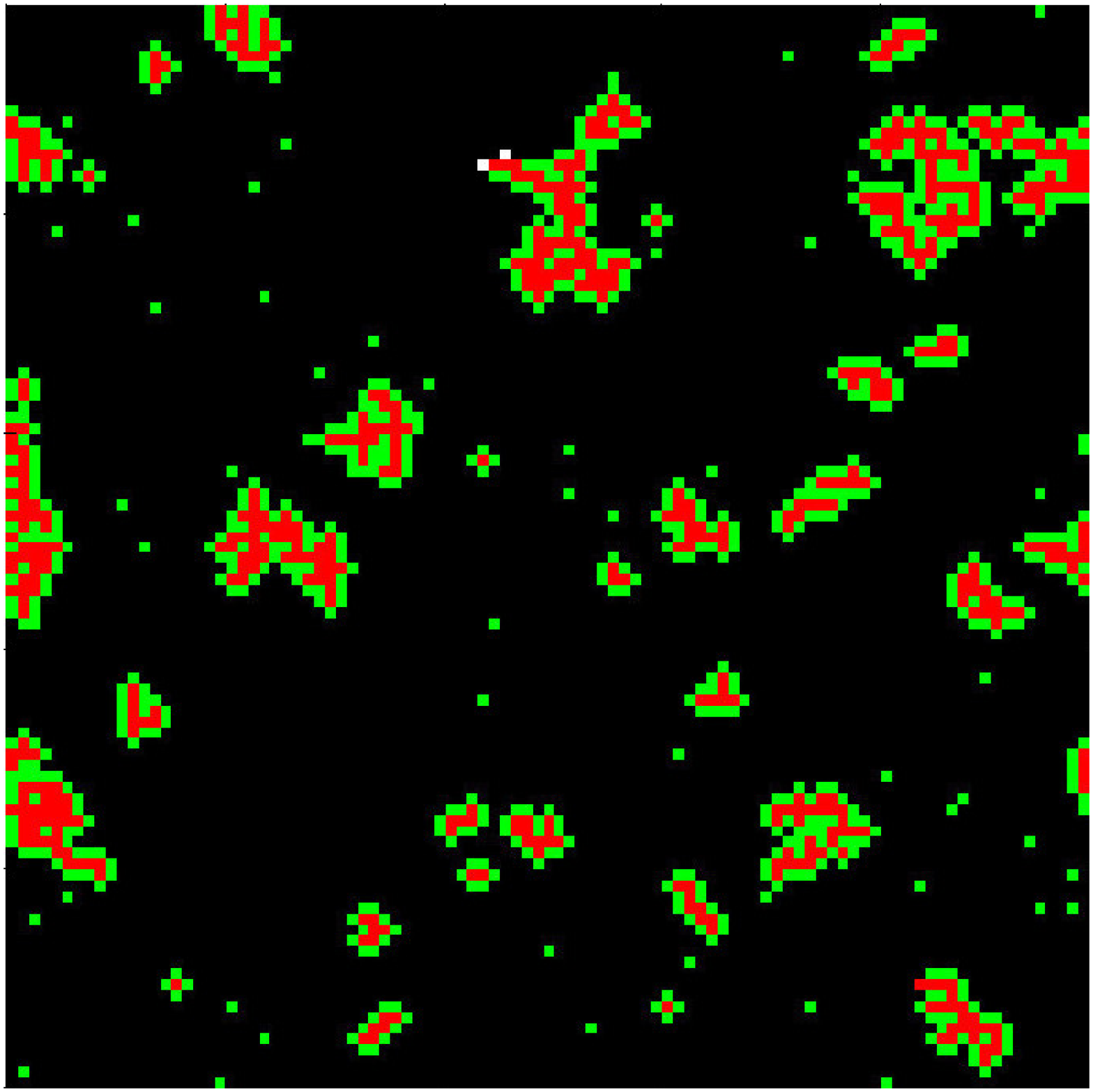}} 
  
  {\includegraphics[width=0.15\textwidth]{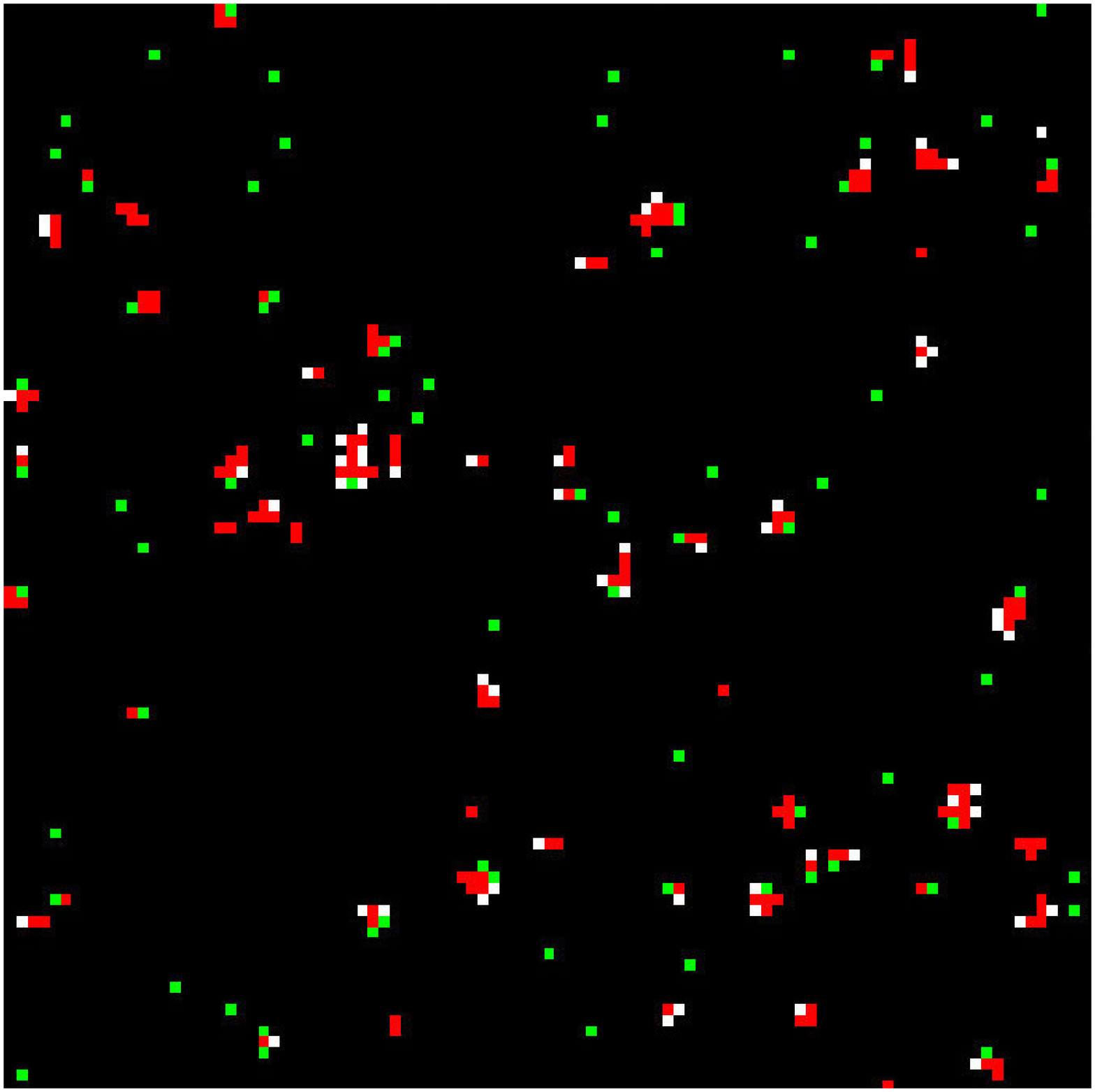}} \hspace{0.5mm}
  {\includegraphics[width=0.15\textwidth]{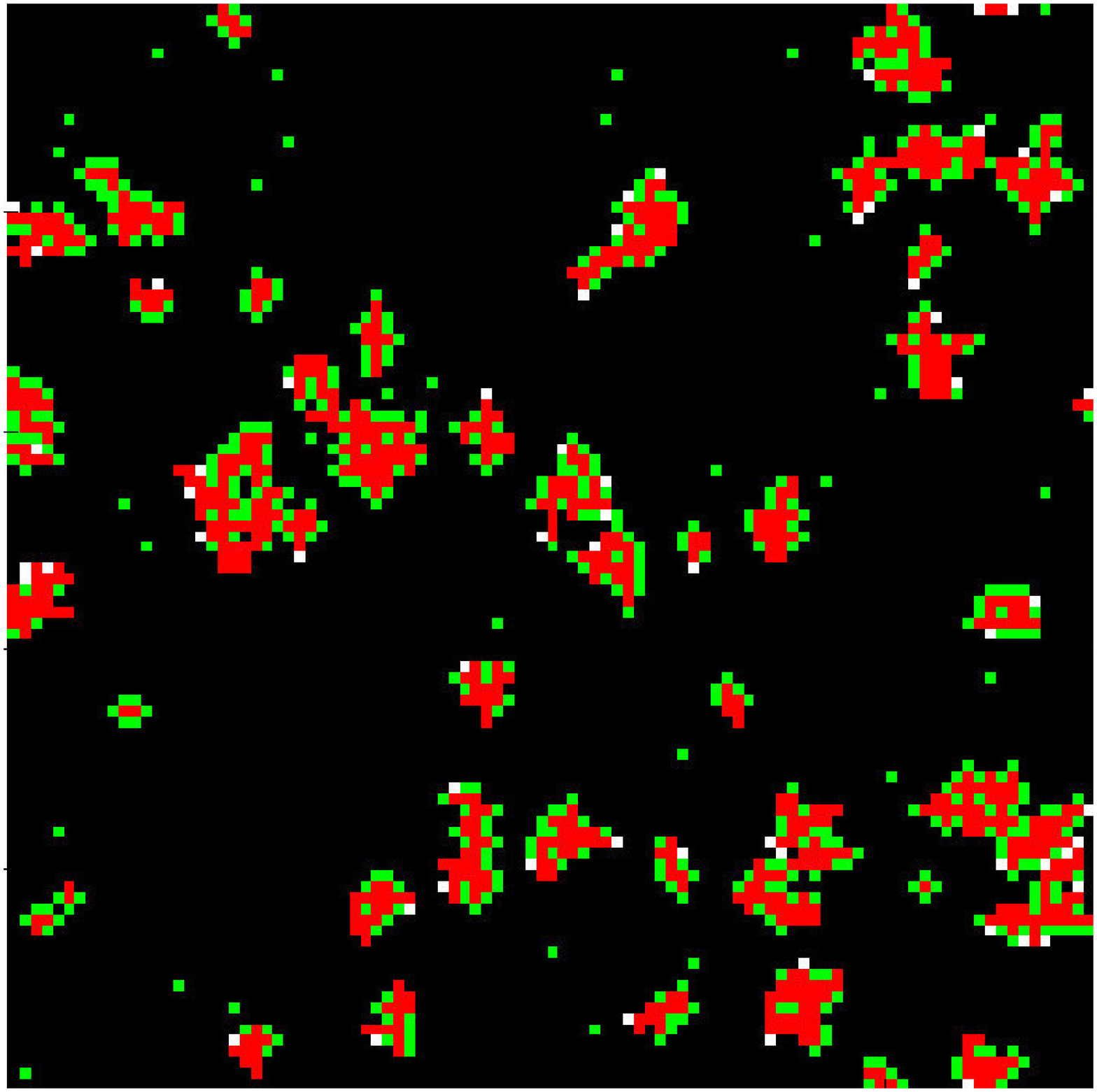}} \hspace{0.5mm}
  {\includegraphics[width=0.15\textwidth]{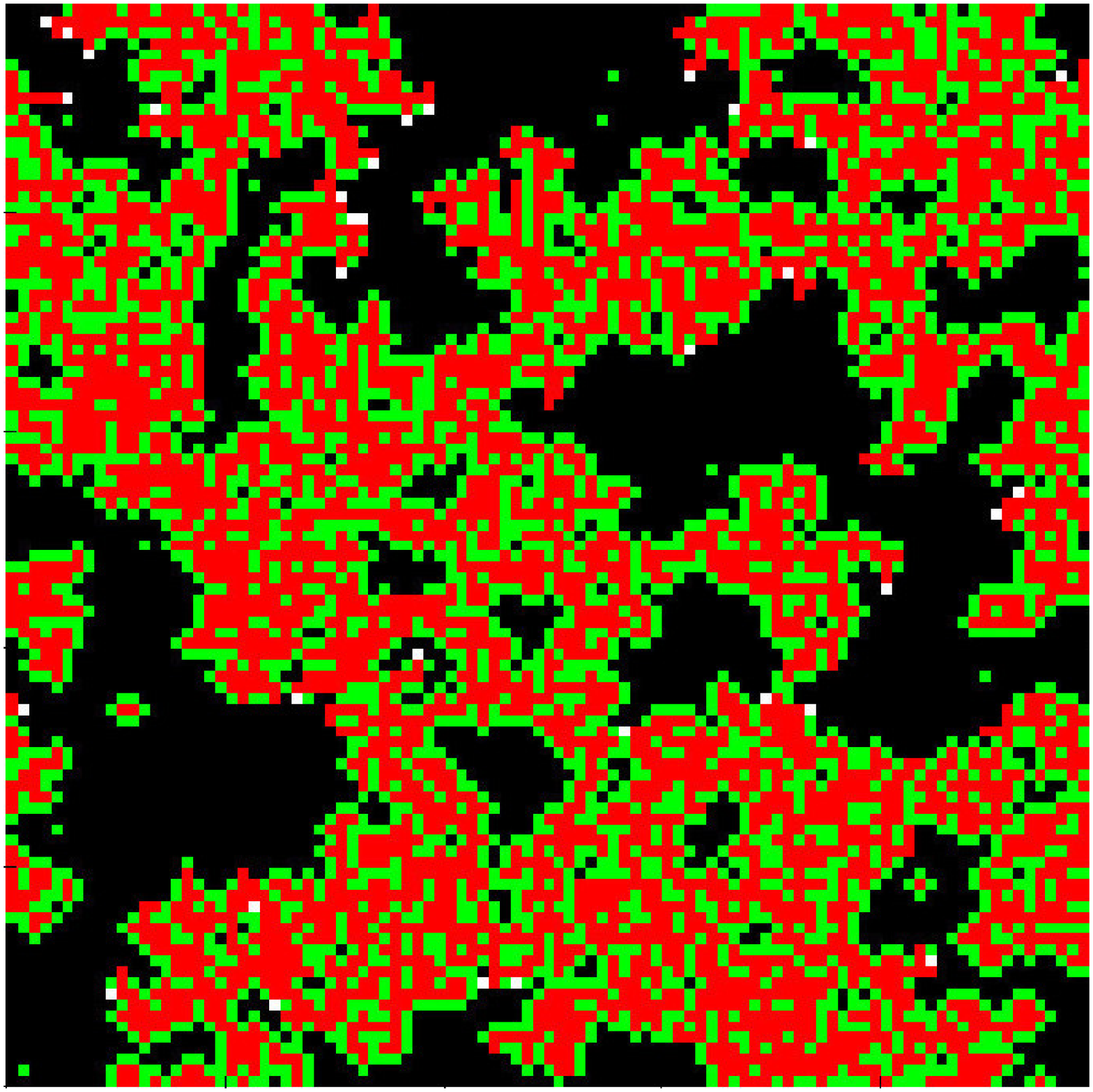}}
  
  {\includegraphics[width=0.15\textwidth]{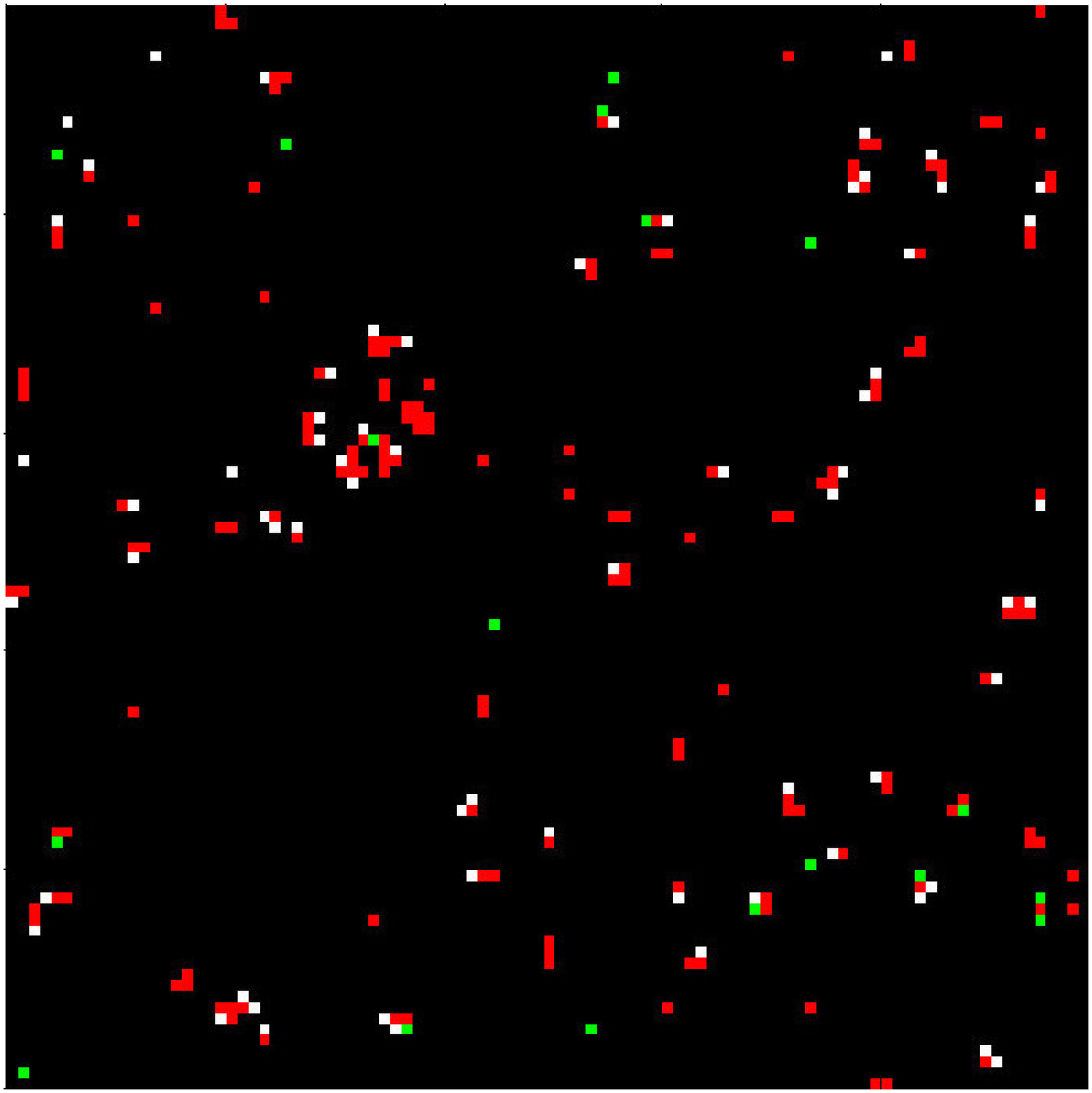}} \hspace{0.5mm}
  {\includegraphics[width=0.15\textwidth]{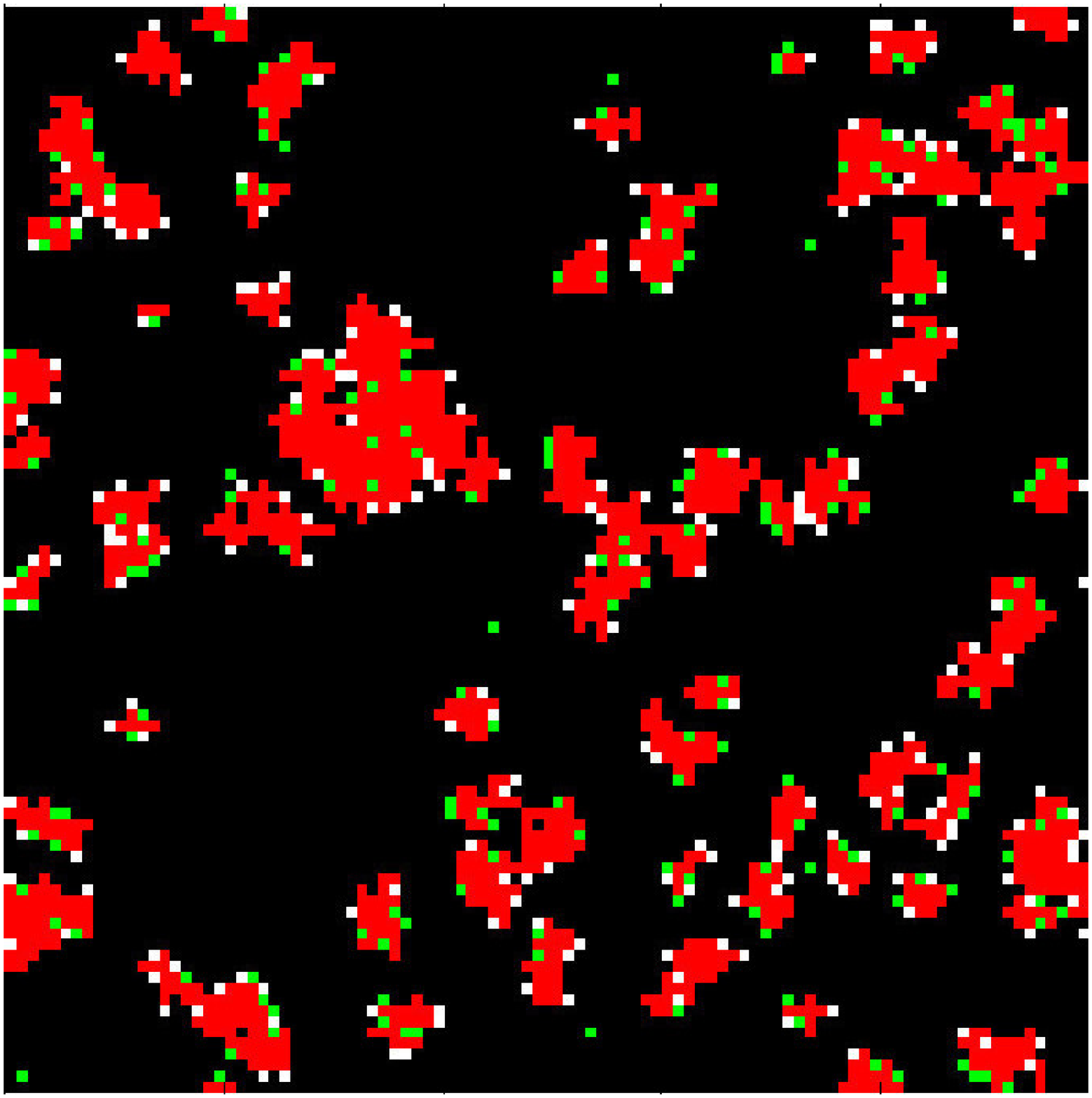}} \hspace{0.5mm}
  {\includegraphics[width=0.15\textwidth]{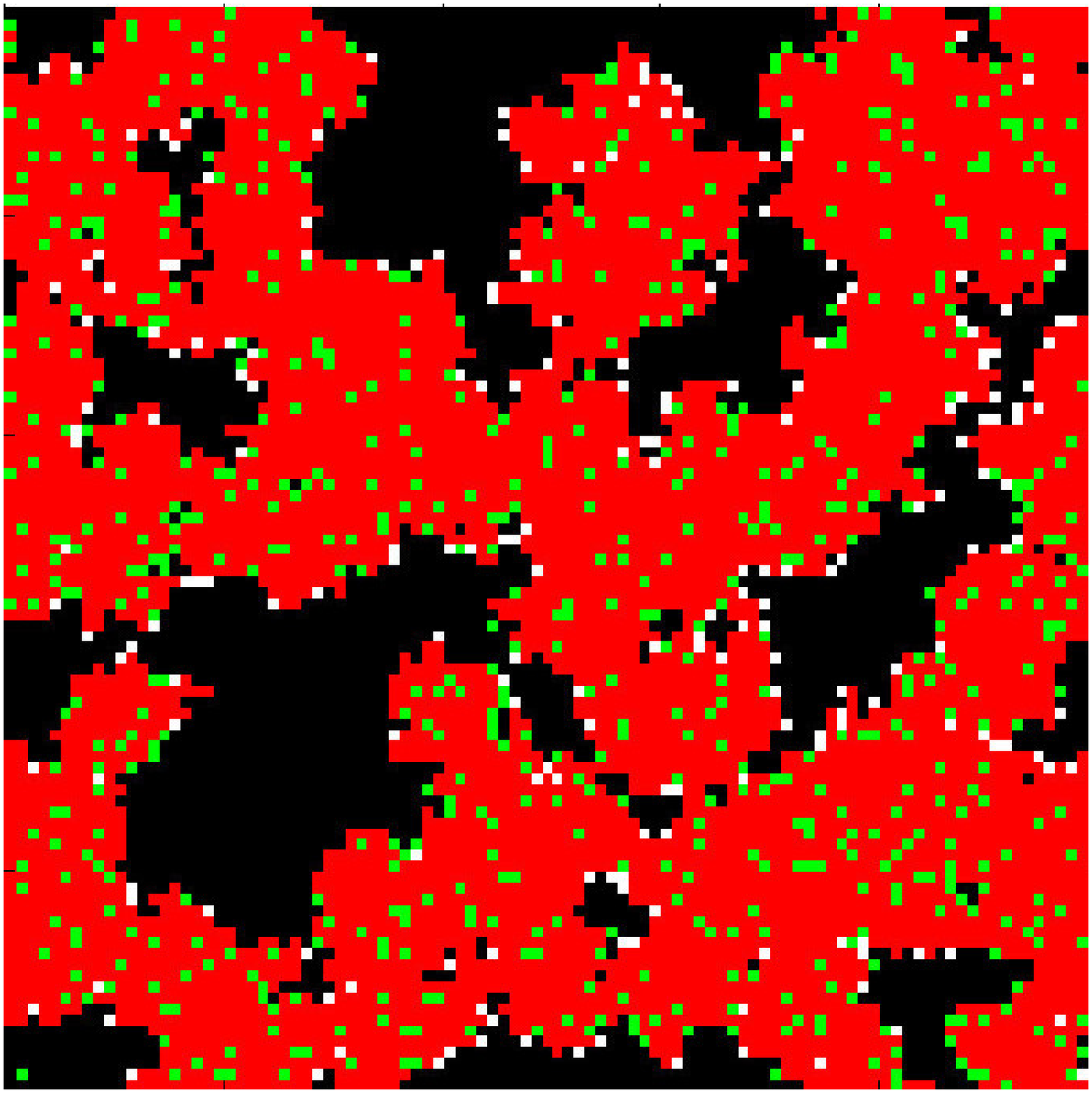}}
  \caption{Snapshots for the square lattice with $\gamma=0.4$, 0.6 and 0.9 (from top to bottom, corresponding to the region before, at and after the peak of $\rhoR$ at $\gamma\simeq 0.6$ in Fig.~\ref{fig.SZ-gamma}) at three different times (increasing from left to right). The color code is: black ($S$), white ($E$), red ($Z$) and green ($R$). Notice that, in the third row, the final snapshot is not the 
asymptotic state of the population (spreaders will dominate and only a few susceptibles survive, isolated by removed agents).}
\label{fig.snapshots}
\end{figure}

Fig.~\ref{fig.snapshots} provides a geometric perspective on the time evolution for the square lattice in the region $\beta\gg\kappa$ and three different values of $\gamma$ around the peak at $\gamma\simeq 0.6$ (see Fig.~\ref{fig.SZ-gamma}). The screening effect of the removed agents strongly changes the evolution of the system. For a value of $\gamma$ to the left of the peak (top row), spreaders remain confined in small groups, despite the large virality ($\beta\gg\kappa$), with removed individuals on the surface. In this case, the number of removed agents necessary to cover the total surface is small.  On the other hand, to the right of the peak (larger $\gamma$, bottom row), the number of removed is small and, as a consequence, not enough to prevent spreaders from invading the whole population. The infection rate is so high that there are almost no susceptibles in the final state. For intermediate values of $\gamma$, close to the peak (middle row) the number of removed individuals in the central panel is not enough to constraint the whole population of spreaders and a small fraction keeps infecting and removing agents from the population.
We now further explore the conditions that induce a large fraction $\rhoR$ in the parameter space.

%
%

\begin{figure}[h!]
  \centering
\includegraphics[width=\columnwidth]{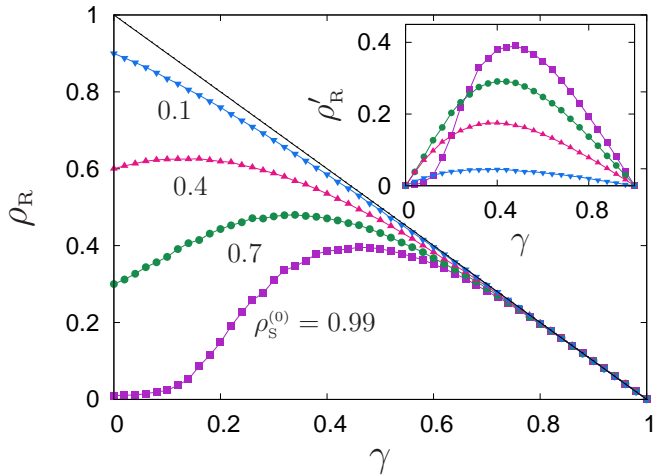}
\caption{Fraction of removed agents $\rhoR$ as a function of $\gamma$ for several values of 
$\rhoSz$ in the scale-free network with $\beta \gg \kappa$. The straight line is the MF result $\rhoR=1-\gamma$. Inset: the corresponding asymptotic fraction of secondary removed individuals after discounting the contribution from the initial state, $\rhoR'=\rhoR-\rhoEz(1-\gamma)$.} 
\label{fig.vagammaR}
\end{figure}

While MF predicts that $\rhoR=1-\gamma$, for the three structured topologies considered here,
the number of removed individuals presents a non-linear behavior as a function of $\gamma$:
there is a minimum value of $\rhoSz$ above which there is a peak in $\rhoR$ when $\beta\gg\kappa$. 
As shown in Fig.~\ref{fig.vagammaR}, for decreasing values of $\rhoSz$, the peak moves to the left, widens and increases in height, eventually arriving at $\gamma=0$ for a finite $\rhoSz$. Obviously, if $\rhoSz=0$, the exposed individuals in the initial state become removed with probability $1-\gamma$, the same result as MF (the upper bound, straight line in Fig.~\ref{fig.vagammaR}). Notice that 
all curves merge with this MF result for large $\gamma$. Even if $\rhoSz>0$,  most of the exposed agents become
spreaders that, in turn (in the limit considered here, of large $\gamma$ and $\beta\gg\kappa$) help remove the remaining
susceptibles. Eventually, removed and spreaders seem to be the only agents remaining (whether susceptibles actually
disappear or a very small fraction remains depends on further, more detailed simulations).
In our model, there are two main mechanisms that remove agents, the reaction $SZ \rightarrow ZR$ with a rate $\kappa$, and the spontaneous decay $E\rightarrow R$ with rate $1-\gamma$.
At the $\rhoR$ peak there are contributions from both processes. The effect of the leading  
mechanism in each side of the peak can be seen in the snapshots of Fig.~\ref{fig.snapshots}.
Thus, the peak  separates the regimes where spreaders still have contact with susceptibles, and the  point where most of the later only survive inside isolated islands surrounded by removed. 
To quantify the removed individuals that do not originate from the exposed agents already present in the initial state, we show $\rhoR'=\rhoR-\rhoEz(1-\gamma)$ in the inset of Fig.~\ref{fig.vagammaR} as a function of $\gamma$. 
Notice that the role played by $\rhoSz$ is reversed when considering only the agents whose removal was inherent to the population dynamics, not those in the initial state: a larger peak is observed when more susceptibles are present in the initial state. Indeed, in the main panel of Fig.~\ref{fig.vagammaR}, the case with the smaller peak becomes the largest in the inset. Moreover, $\rhoR'$ presents a peak for all values of $\rhoSz$, even those for which $\rhoR$ is monotonic.
After the peak, the linear behavior of $\rhoR'$ is a direct consequence of the collapsed behavior seen in the corresponding 
region of Fig.~\ref{fig.SZ-gamma}: in that region, $\rhoR\simeq 1-\gamma$ and $\rhoR'\simeq 1-\gamma - \rhoEz(1-\gamma) = \rhoSz(1-\gamma)$. Indeed, by plotting $\rhoR'/\rhoSz$, a similar collapse is obtained (not shown). 
This can be seen as a decoupling of the effect of $\gamma$ on the final population of $\rhoR$, i.e., the secondary removed individuals will depend on $\gamma$ but not in a simple way.



\section{Conclusions}
\label{section.conclusion}

We considered a simple rumor spreading model that includes agent-level skepticism~\cite{AmAr18}. 
An individual that had contact with a rumor may become removed (disinterested)
from the propagation process following two different routes, both driven by skeptical
inquiry. First, if it is a spreader,
by getting in contact with a skeptical susceptible. This is a direct, active mechanism based
on persuasive argumentation. Second, if it has only being exposed but not yet turned into a spreader, 
by discarding the rumor (passive mechanism) 
due, for instance, to fact-checking. Prior work~\cite{AmAr18} studied this model within the 
mean-field approximation (MF). Here we extend those results by 
including correlations among neighboring pairs of agents (pair approximation, PA), and compare with 
agent-based simulations in different geometries (one- and two-dimensional lattices and random 
and scale-free networks), showing that the two-sites approximation well describes the simulation 
results. Neighborhood, in this case, refers to social contacts as spatially close individuals
not always exchange information.


Removed sites have an important function acting as barriers for rumor spreading when spatial effects 
are taken into account. In the asymptotic, stationary state, these removed agents may coexist with susceptibles 
and spreaders. Mean-field, being equivalent to a fully-connected system, does not provide
such possibility, and either S or Z get extinct (preliminary results show that when including diffusion, as the 
effective range of interactions increase, the lattice results approach those of MF). Notice also that although
these MF solutions were not observed in the simulations, more extensive simulations should be performed in order
to completely rule this possibility out. 
In order to be stable, the coexisting susceptibles and spreaders must 
be spatially isolated, their interactions hindered by the surrounding removed individuals. 
Square lattices may present a larger fraction of susceptible individuals as compared to complex networks for a given set 
of parameters. The later has, instead, more removed agents (a relevant parameter as it corresponds to those
that lost interest in spreading the rumor). The reason is that susceptibles form more compact groups in regular lattices, thus
needing a smaller number of protective removed agents on their surface. The long-range interactions present in the
complex networks increase the probability of SZ encounters, and a correspondent larger number of removed individuals.
This indeed is the mechanism allowing social-networks to easily spread information. Interestingly collective, long-ranged conveyors of information like magazines, radio and television, present before the advent of internet, did not have such capability because of the traditional fact-checking that most of the media enforces. When individuals start to propagate their own beliefs and opinions, because skeptical inquiry and critical thinking are not, yet, widely held capabilities, rumor contention becomes a very difficult task to which,
presently, there is no efficient solution available. 

Some spatial contention may be obtained by attaining a 
sufficient degree of herd immunity. A similar result, for disease spreading, is traditionally obtained
through vaccination programs. Although not all individuals may have their skeptical immunological system fully 
developed, pseudosciences and fake-news propagation may be halted by protecting 
vulnerable groups through a large and well distributed population of sufficiently  educated people.
It is important to stress again that, in the stationary state, spreaders and removed individuals may coexist. As is indicated by
experimental observations, while new conspiracy theories and pseudosciences are constantly invented and disseminated, they rarely replace
entirely the previous ones. Not even when actively targeted by skeptical individuals. In these situations, while completely eliminating 
fake news spreaders may not be feasible, achieving the optimal fraction of individuals not interested in rumors, i.e., removing
then from the propagation process, can be a more realistic goal.
We observe that intermediate values of $\gamma$ (the rate of spontaneous transitions between
exposed and spreader) seem to generate a larger fraction of removed individuals when the rumor is highly viral ($\beta\gg\kappa$). Interestingly, after correcting the final fraction of removed to exclude the initially exposed individuals that will unavoidably be removed, we observe that this effect is even more pronounced when the initial number of exposed individuals is small.  

From the perspective of fake-news spreading, this can be understood as a protective effect that an initial exposure can cause in a (sufficiently) skeptic population. If there are no passively skeptic individuals (high $\gamma$), the rumor will, evidently, dominate the population, with a high final $\rhoZ$. But if the passive skepticism is too high (low $\gamma$), the rumor will be quickly trapped inside a small island of removed (disinterested) individuals. While this can create a large final fraction of susceptible agents, it also means that most individuals never had contact with the rumor, leaving the final population still susceptible to future rumors. Finally, an intermediate value of $\gamma$ near this point can make the rumor spread through the population in a controlled way, creating more removed individuals than any other case. While this is not a perfect scenario, a population with the highest number of disinterested individuals will be more resilient to rumor spreading.

Ref.~\cite{AmAr18} emphasized that the mechanism used by skeptical agents to individually 
counteract the propagation of rumors 
is similar to the pop culture scenarios for a zombie outbreak (both rumors and zombies may be directly eliminated by susceptibles agents). 
Such picture has been previously used to communicate the science of real epidemics and the advantages of
preparedness~\cite{CDC}, 
while also motivating some theoretical studies~\cite{MuHuImSm09,AlBiMySe15,HoWa20}. 
 On a global level, the current dissemination
of pseudosciences and disinformation, along with the widespread phenomenon of authorities discrediting scientists (e.g., the
recent climate emergency and coronavirus epidemic), everything facilitated by our technology, is the equivalent of a zombie
apocalypse. Thus, this class of models remains interesting as we may get insight on how skeptics 
should act to stop rumor spreading. In particular, a better understanding of how rumor propagates 
and how effective can be an effort to resist or even to demove people may help to devise
intervention strategies focused on specific individuals in a similar
way as crime and internet hate control.


\begin{acknowledgments}
WGD acknowledges the hospitality of the IF-UFRGS during his stay where part of this work was done.
Work partially supported by the Brazilian agencies FAPERGS, FAPERJ, CNPq (process number 428653/2018-9), and CAPES (Finance code 001). 
\end{acknowledgments}


\end{document}